# Acceleration of shell DFT-1/2 in high-throughput calculations via cutoff radii prediction

Shanzhong Xie, Kan-Hao Xue,* Zijian Zhou, Xiangshui Miao

School of Integrated Circuits, Huazhong University of Science and Technology, Wuhan 430074, China

*Corresponding author, email: xkh@hust.edu.cn

**Abstract**

Shell DFT-1/2 is a fast band gap rectification method that is versatile for semiconductor supercell and superlattice calculations, which involves two cutoff radii that have to be optimized. Although such optimization is trivial in terms of time cost for a primitive cell, in high-throughput calculations this can be a big concern because most materials are themselves in small unit cells. The numerous optimization trials increase the computational cost to orders of magnitudes higher. In this work, we construct a regression model for the prediction of the two cutoff radii based on chemical composition and primitive cell structure. Moreover, a model for metal and insulator classification is also given, with 95.2% accuracy.

The band gap ($E_g$) is an ordinary but extremely important parameter in characterizing semiconductors and insulators.[1] Although feasible in principle, measurement of the band gap suffers from experimental cost and the difficulty in material synthesis, hindering the high-throughput acquirement. With the emergence of high-throughput calculation technique using typically density functional theory (DFT), the process of material screening has been significantly accelerated.[2–5] Yet, the solid-state electronic structure calculation using DFT is not a straightforward task.[6,7] For example, under the most widely implemented functional of the Perdew-Burke-Ernzerhof (PBE) form, which has the generalized gradient approximation (GGA) flavor, the calculated band gaps are systematically lower than experimental, by ~40%. Beyond conventional DFT at the GGA or local density approximation (LDA) levels, methods such as hybrid functionals[8,9] and $GW$-type[10,11] quasi-particle methods could significantly improve the quality of $E_g$. Nevertheless, these methods are currently not yet suitable for high-throughput computation due to their prohibitively high computational costs.

The DFT-1/2 method as proposed by Ferreira, Marques and Teles in 2008,[12,13] and its variant—the shell DFT-1/2 (shDFT-1/2) method[14–17] as proposed in 2018, provide an alternate route that only uses the self-energy potentials to correct the spurious electron self-interaction due to LDA or GGA. In particular, shDFT-1/2 could well recover the electronic band structure of Ge,[14] without referring to any empirical parameter, and it shows excellent band gap accuracy in Sb-based semiconductor superlattices for infrared-detection.[18] The computational cost of shDFT-1/2 is merely at the LDA/GGA level, and in certain cases it even outperforms LDA/GGA in terms of speed.[15] This renders the shDFT-1/2 method a highly promising candidate for high-throughput calculations in material genome engineering, as it effectively balances computational efficiency with accurate band gap predictions for most semiconductors and insulators. However, a notable challenge arises from the fact that shDFT-1/2 requires the determination of two cutoff radii for the self-energy potential—namely, the outer cutoff radius ($r_{out}$) and the inner cutoff radius ($r_{in}$)—unlike conventional DFT-1/2, which only involves $r_{out}$. These radii must be determined variationally, typically through a comprehensive scanning process that ideally involves a two-dimensional (2D) joint optimization of $r_{in}$ and $r_{out}$. This is not a big concern in large supercell calculations, because one derives the values of $r_{in}$ and $r_{out}$ in small primitive cells, which are directly transferable to large supercells, but in high-throughput calculations for small cells this adds to the computational complexity greatly. Indeed, optimization of these radii costs much more time than the desired shDFT-1/2 product run for a typical small primitive cell. While shDFT-1/2 has found widespread application in large supercell calculations, such as the study of defects,[19,20] elemental substitution,[21] and interfaces,[22] its adoption in material genome engineering remains in its infancy.

In this work, we strive to improve the efficiency for the cutoff radii optimization of shDFT-1/2 in terms of physical models and machine learning (ML). In shDFT-1/2, a shell-like trimming function is used to confine the spatial range of the self-energy potential

$$\Theta(r) = \begin{cases} 0 & r < r_{in} \\ \left\{1 - \left[\frac{2(r - r_{in})}{r_{out} - r_{in}} - 1\right]^p\right\}^3 & r_{in} \leq r \leq r_{out} \\ 0 & r \geq r_{out} \end{cases}$$

where $p$ is an even integer of power index, which should be sufficiently large and is recommended to be 20 by default (the original DFT-1/2 uses a smaller $p = 8$ since it does not need to trim the potential at the near-core region). To fix $r_{in}$ and $r_{out}$, one should scan them as to maximize the band gap. After trimming, the self-energy potential will be attached to the pseudopotential of the anion for the self-energy correction to the valence band. To ensure the correct combination of $r_{in}$ and $r_{out}$ for any specific element, the 2D optimization requires a great number of self-consistent field (SCF) calculations. For instance, if one selects the region 0 to 5.0 Bohr for both $r_{in}$ and $r_{out}$, with a step of 0.1 Bohr, the total number of trial SCF calculations would be nearly 50 × 50 = 2500. In the case of compounds where two elements require self-energy correction, this computational cost becomes even more pronounced. Hence, one may adopt another optimization scheme, as to first optimize $r_{out}$ with $r_{in}$ fixed to zero, and then optimize $r_{in}$ using the $r_{out}$ outcome. Subsequently, one should further optimize $r_{out}$ using the as-optimized $r_{in}$ value. The procedure continues until the optimal $r_{in}$ and $r_{out}$ no longer change. This scheme is in general more efficient than the 2D optimization, but the first two steps still involve searching within a vast scope. Hence, the algorithm should be further improved, with an initial prediction of $r_{in}$ and $r_{out}$ highly desirable.

ML methods have gained increasing applications in condensed matter physics and materials science due to their cost-effective nature, serving as a powerful tool now to predict electronic properties of new materials through integration and mining of large-scale material databases.[23,24] Current research efforts have explored ML models for the task of semiconductor band gap prediction, but these approaches predominantly rely on data labels derived from PBE calculations, experimental measurements, or *GW* methods, which suffer from either limited accuracy or prohibitive acquisition costs. For instance, Zhou *et al.* developed models using experimental band gaps as the dataset resource.[25] Lee *et al.* constructed a dataset of 270 inorganic compounds with $G_0W_0$ calculation results as data labels.[26] Wu *et al.* employed hybrid datasets, through combining experimental band gaps with PBE-predicted values.[27] In our work, to address any existing limitations, we plan to integrate ML methods with the shDFT-1/2 calculation data. The objective of our approach is to predict the two critical radii ($r_{in}$ and $r_{out}$), such that the following shDFT-1/2 product run could be accelerated using ML-predicted initial radii, thus enabling refined band gap calculations in an efficient manner. And the paper is organized as follows. We first introduce the global workflow of our approach, and the metal/insulator classification model will be derived first. Using a similar strategy, two models for predicting the shDFT-1/2 inner cutoff radius and outer cutoff radius for self-energy potential will be proposed.

The overall workflow of this study is illustrated in **Figure 1**. We constructed two independent datasets to train a cascaded modeling system consisting of a metal/non-metal classification model (MC model for short) and a radii regression model (RR model for short) for predicting the two radius parameters in shDFT-1/2. We focus on self-energy correction on a single element of a binary compound material, thus the exact way of implementation can be called shDFT-0-1/2.[14] In case both elements of the binary compound require self-energy correction, shDFT-1/4-1/4 ought to be carried out instead. In that case, one can simply use our method to predict the initial cutoff radii for each element separately. For this reason, our dataset only involves binary compounds that are subject to shDFT-0-1/2 correction, and the current focus is therefore more focused on the electronegative elements. Our implementation begins with "feature engineering"[28,29] to eliminate redundant information through dimensionality reduction techniques. Accordingly, the relationship between structure and target property could be established. Hyperparameter optimization is performed using a grid search across pre-defined parameter spaces, with model robustness subsequently validated through 5-fold cross-validation. Upon completing model training, the inference pipeline operates in several sequential stages. (i) Structural parameters are first fed into the system. (ii) The MC model classifies the material as a metal or a non-metal, where only non-metals are being subject to self-energy correction. Metals, on the other hand, will be handled by conventional GGA calculations. (iii) The RR model predicts optimal cutoff radii that will be useful for the subsequent shDFT-1/2 calculations, which limits the scope of searching and diminishes the number of trials in optimizing $r_{in}$ and $r_{out}$. (iv) The final shDFT-1/2 run with the optimized cutoff radii yields the rectified electronic structure.

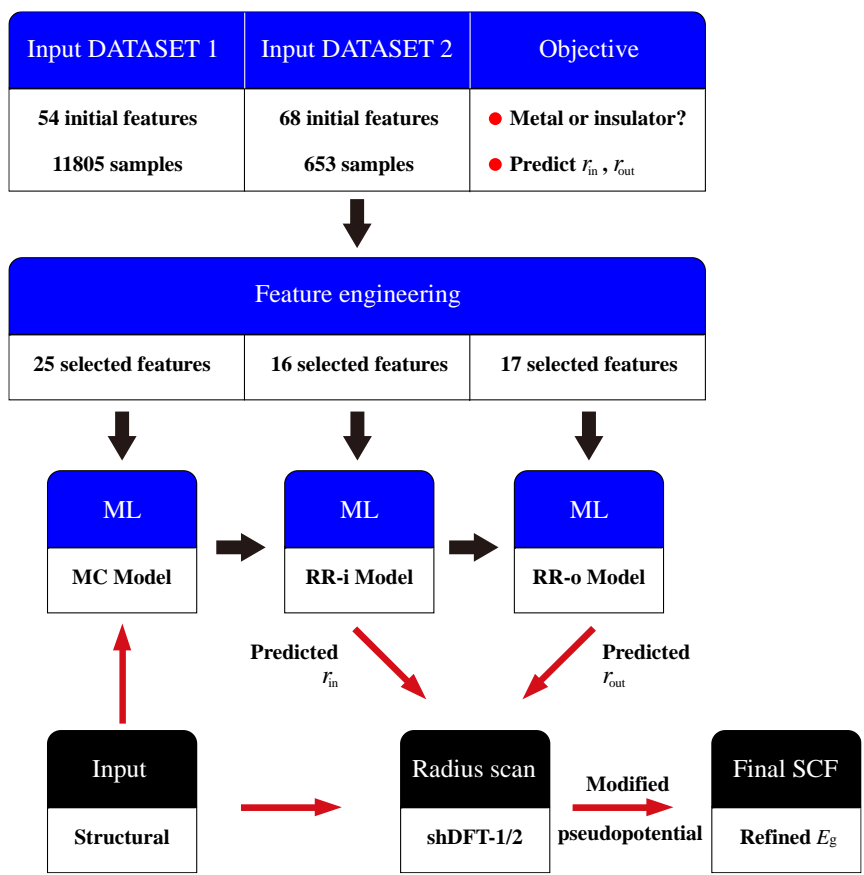

**Figure 1.** A schematic of the model training process and the inference framework. The blue boxes represent the steps involved in model training, while the black boxes illustrates the process of using the model for inference.

Algorithm selection is also critical in model performance optimization for multi-class classification tasks. To systematically evaluate the relative efficacies of various learning paradigms, our training framework incorporates 8 distinct models, including decision trees (DT),[30] gradient boosting classifier (GBC),[31] K-nearest neighbors (KNN),[32] logistic regression (LR),[33] naive Bayesian (NB),[34] as well as support vector machines (SVM)[35] that permit three distinct kernel implementations (linear, poly, and radial basis function (RBF), where we only select linear and RBF).[36] These algorithms are in fact far from exhaustive, but they have been strategically selected to include representative approaches spanning rule-based systems, ensemble methods, instance-based learning, probabilistic classifiers, and maximum-margin classifiers.

The complete input features, corresponding labels, and their reference sources for DATASET D1 are specified in **Supplementary Note 1**, containing 11805 initial samples and 54 initial features. The most straightforward information for the prediction tasks is the chemical composition of the material. However, the incorporation of both elemental composition descriptors and crystallographic structural characteristics seems to be more effective. Subsequent feature engineering addressed the "curse of dimensionality" risk. The initial 54 features were subject to Pearson correlation coefficient (PCC) filtering. Provided that a few features share a correlation greater than 0.9, several members of this sub-set may be eliminated. Through the max-relevance and min-redundancy (mRMR) selection, it can be decided which one or few members should be retained within this sub-set. This procedure successfully identified 41 relevant descriptors out of 54, but the extent of optimization is still insufficient. A "last-place elimination" methodology as proposed by Lu *et al.*[37] was further used to optimize the number of features. The method was initially designed for regression, and we had to extend it to classification tasks. Hence, the GBC algorithm, which is suitable for feature importance (FI) evaluation, was used to compute averaged FI in our case. In the first round of training, all 41 features were input, and the GBC algorithm was implemented for 100 times to obtain the average FI values. And the average F1-score of this 41-feature model was calculated as well. The least important feature was then eliminated, leaving 40 features. These features were used for the 2nd round of training, which yielded the updated average FI values, indicating the second eliminable feature. The process was carried out over and over again, until only one feature remained. That feature turned out to be the most significant one. The last but one round, therefore, retained this feature as well as the second most significant one. Hence, one could trace back to pick out the desired number ($M$) of significant features. To obtain the optimal choice of $M$, one should inspect the value of F1-score in each round, as well as AUC (area under the curve) performance. In general, the $M$ value leading to the maximum F1-score should be selected, but in case several $M$ values share similar F1-scores, the AUC parameter could also be useful. After such "last-place elimination" process, only 25 descriptors were retained. In our case, $M = 25$ corresponds to the highest F1-score of 0.9362 as well as the highest AUC value (0.9484), as shown in **Figure S1**. Notably, some features do demonstrate crystallographic relevance rather than elemental composition dependence, such as "min_distance" (the distance to the nearest neighbor atom) and "max_distance" (the distance to the farthest atom within the coordination).

**Table 1.** A performance comparison of several classifiers based on various evaluation metrics: accuracy, AUC, and precision ratios for both CLASS-0 (non-metals) and CLASS-1 (metals).

| Classifier | Accuracy | AUC | Pre-0 | Pre-1 |
| --- | --- | --- | --- | --- |
| DT | 0.920 | 0.920 | 0.90 | 0.94 |
| GBC | 0.952 | 0.991 | 0.94 | 0.97 |
| KNN | 0.939 | 0.939 | 0.91 | 0.97 |
| LR | 0.820 | 0.881 | 0.80 | 0.85 |
| NB | 0.788 | 0.851 | 0.79 | 0.78 |
| SVM (linear kernel) | 0.807 | 0.876 | 0.81 | 0.81 |
| SVM (RBF kernel) | 0.947 | 0.983 | 0.93 | 0.97 |

There is one issue in DATASET D1, that it only involves 2632 non-metallic samples, but up to 9173 metallic samples. To mitigate the class imbalance problem, we utilized the Borderline-SMOTE[38] method to synthesize non-metallic specimens theoretically. One may also choose to synthesize the specimens before feature selection, but a previous research[39] demonstrates an enhanced model accuracy following the sequence we adopted (feature selection first). In this way, we obtained an entirely balanced sample set, with 9173 metals and 9173 non-metals. The dataset was divided into two parts, 80% for training and 20% for testing. For the purpose of training (using 8 possible paradigms), each paradigm had a hyperparameter optimization process. Via a grid search with 5-fold cross-validation, GBC was identified as the most superior algorithm among the eight. As shown in **Table 1**, the precision metrics for non-metallic (Pre-0) and metallic (Pre-1) classifications demonstrate comparable values across all models, statistically validating the balanced efficacy of the Borderline-SMOTE implementation. In our work, the 0.952 accuracy and 0.991 AUC value outperform a previous composition-only model (0.92 accuracy and 0.97 AUC),[25] and are comparable to DFT-based classifiers (typical reported AUC values are 0.98[40] and 0.93[41]).

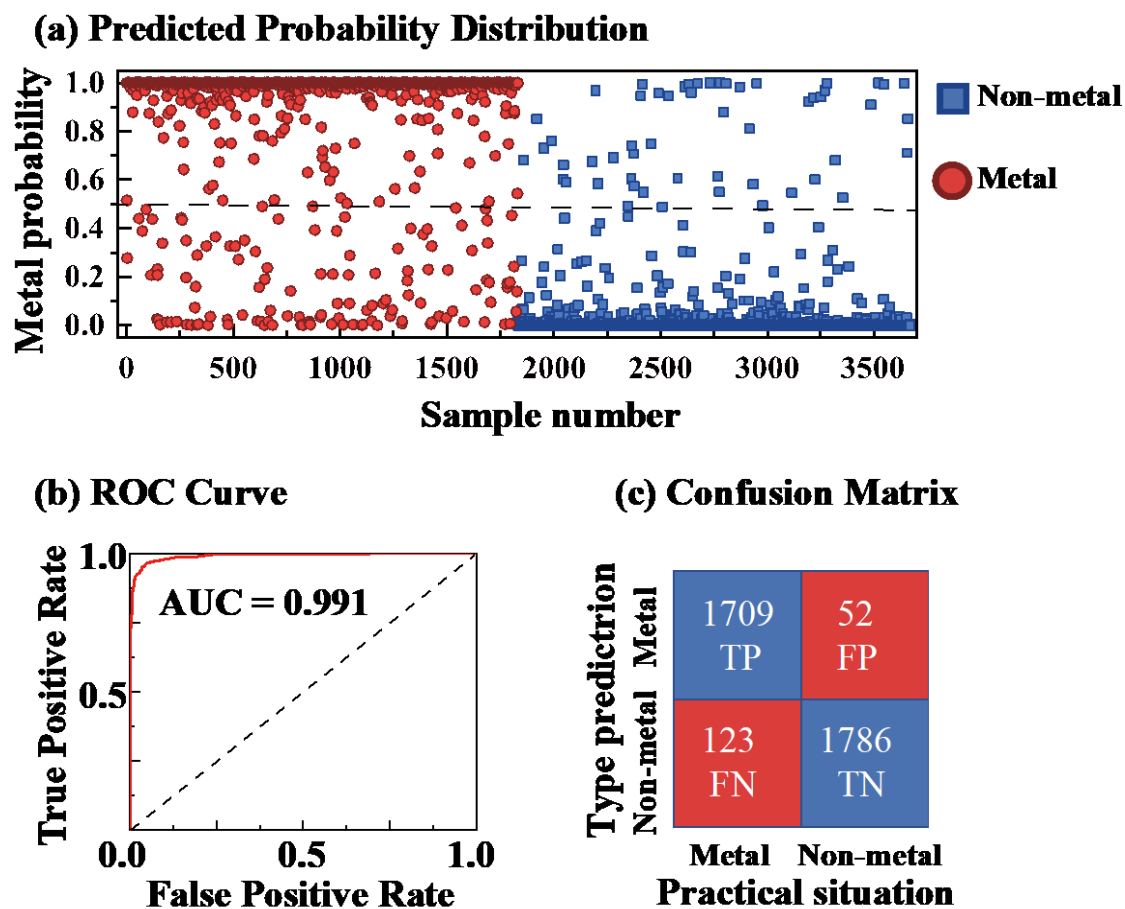

**Figure 2.** (a) The predicted metal probability for each sample, with red circles representing metal samples and blue squares representing non-metal samples. (b) The ROC curve showing the performance of the model. (c) Confusion matrix summarizing the quality of model prediction, where blue and red colors stand for correct and incorrect inferences, respectively.

We then employed the GBC algorithm to obtain the probability distribution for metallicity. For each composition in the test set (3670 samples in total, with half being metallic and half being nonmetallic), the predicted probability of being classified as a metal (0.5-1) or non-metal (0-0.5) was calculated. As shown in **Figure 2a**, the majority of compositions demonstrate clear class segregation with minimal overlap across the probability spectra. Specifically, only 4.7% compounds (175 out of 3670) in the test set were misclassified. Further validation through examination of the ROC curve (**Figure 2b**) and confusion matrix (**Figure 2c**) confirms the robust capability of the model.

For any compound predicted as a non-metal, the self-energy corrected shDFT-1/2 calculation ought to be carried out for the determination of its electronic structure. Therefore, a prediction scheme for the initial values of $r_{in}$ and $r_{out}$ is required to accelerate the radius scanning procedure. This is in particular feasible using the ML methodology. In materials science research, several supervised ML regression algorithms, such as gradient boosting regression (GBR)[42] and artificial neural networks[43,44] have been successfully employed in various applications. In our study, we implemented six distinct ML regression algorithms, including kernel ridge regression (KRR), Gaussian process regression (GPR), support vector regression (SVR), decision tree regression (DTR), GBR, and multilayer perceptron (MLP). Their performances will be subject to a thorough comparison.

To proceed, feature engineering is still the primary concern. The complete input features, label information, and reference sources of DATASET D2 are detailed in **Supplementary Note 1**, containing 653 initial samples and 68 initial features. For the inner and outer cutoff radii parameters in D2, a comprehensive data acquisition was first conducted. To obtain high-quality labels, we exhaustively explored all possible combinations of $r_{in}$ and $r_{out}$. Specifically, the value ranges for both radii were set from 0 to 5.0 Bohr, with a step size of 0.1 Bohr. And a SCF calculation was performed for each combination. Although this strategy incurred an extremely high computational cost, it ensured label accuracy. On account of this precision-induced complexity, the following criteria were established for material screening—number of atoms in a primitive cell no greater than 10, and three-dimensional non-metals only.

In addition, we observed a useful feature, that the sensitivity of shDFT-1/2 band gap with respect to $r_{in}$ is less obvious than that of $r_{out}$. Based on this phenomenon, we choose to predict $r_{in}$ first, which can then be used as an input to predict $r_{out}$. The benefit lies in that $r_{out}$ is practically more important, thus its prediction with a fairly good $r_{in}$ will be more faithful. Hence, we set up two models designated as RR-i (inner radius regression model) and RR-o (outer radius regression model), where the outcome of RR-i serves as an input parameter for RR-o.

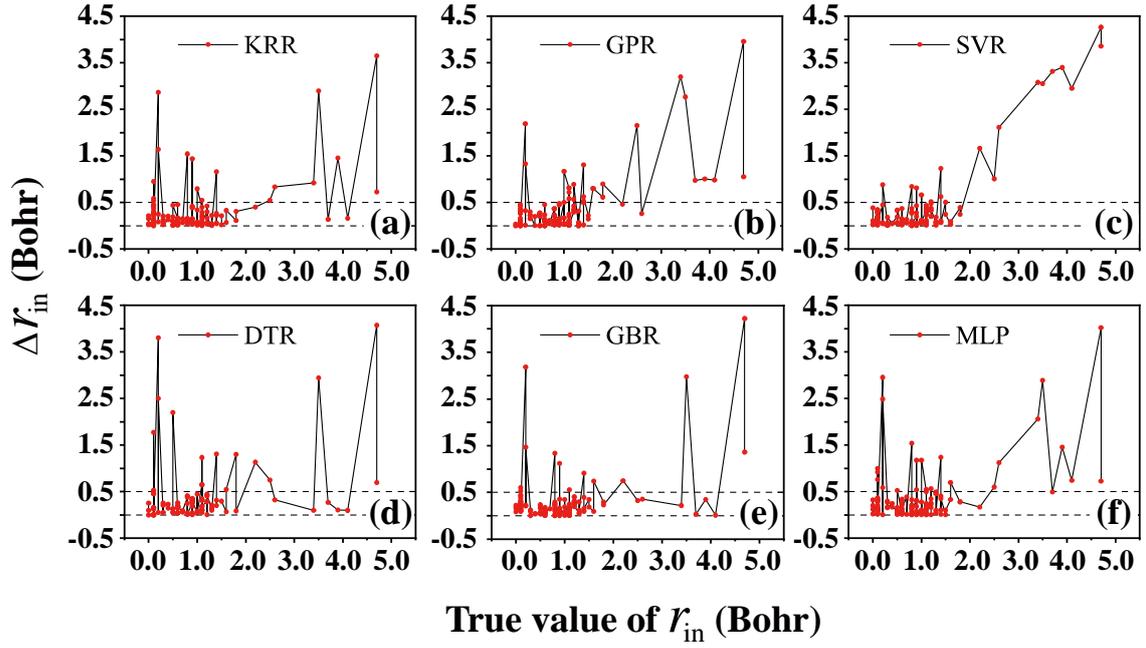

**Figure 3.** Prediction error ($\Delta r_{in}$) versus true inner radius ($r_{in}$) for the six regression models. (a) KRR; (b) GPR; (c) SVR; (d) DTR; (e) GBR; (f) MLP. The two horizontal dashed lines and the bottom (at vertical coordinate -0.5) solid line emphasize the ±0.5 threshold used in the RAE evaluation.

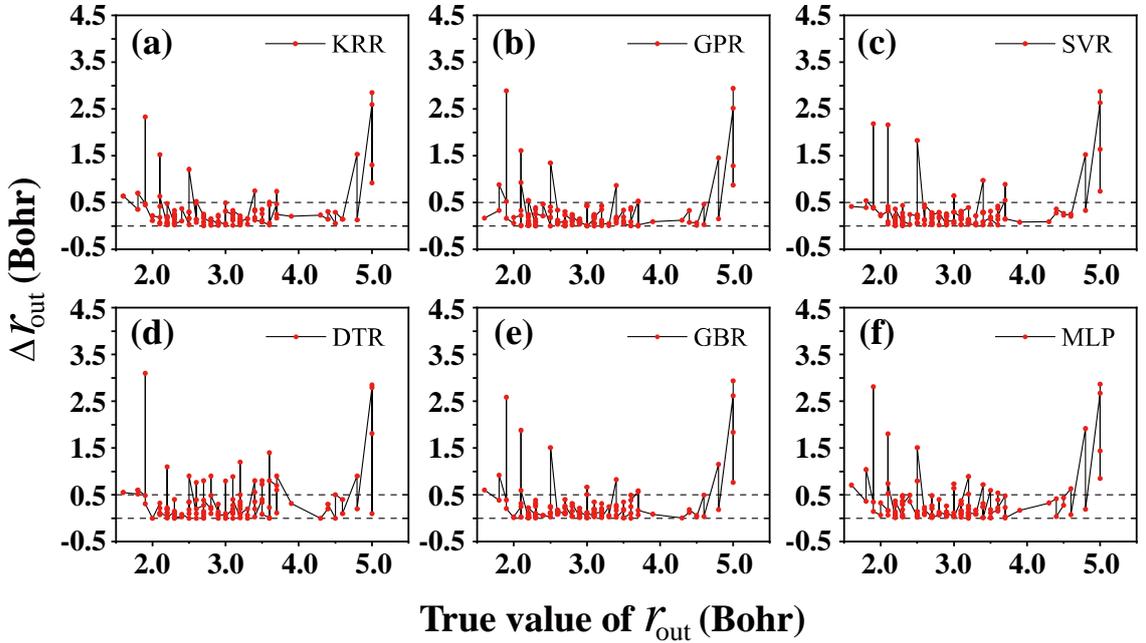

**Figure 4.** Prediction error ($\Delta r_{out}$) versus true outer radius ($r_{out}$) for the six regression models. (a) KRR; (b) GPR; (c) SVR; (d) DTR; (e) GBR; (f) MLP. The two horizontal dashed lines and the bottom (at vertical coordinate -0.5) solid line emphasize the ±0.5 threshold used in the RAE evaluation.

**Table 2.** Performance comparison of different regression models based on the $RAE_\alpha$ metric for RR-i and RR-o.

| Regressor | $RAE_\alpha$-RR-i | $RAE_\alpha$-RR-o |
|:---:|:---:|:---:|
| KRR | 0.87 | 0.89 |
| GPR | 0.82 | 0.89 |
| SVR | 0.86 | 0.90 |
| DTR | 0.86 | 0.81 |
| GBR | 0.90 | 0.88 |
| MLP | 0.82 | 0.84 |

We further define a performance evaluation metric, $RAE_\alpha$ (radius accuracy evaluation parameter using threshold $\alpha$) as follows. Any prediction is being considered correct if the predicted radius falls within the $\pm\alpha$ range of the true radius, and the true radius should come from a thorough 2D optimization of both $r_{in}$ and $r_{out}$. Provided that the predicted radius is outside this range, it is considered an incorrect prediction. We assess the model performance by calculating the ratio of the number of correct predictions to the total number of test samples. For both RR-i and RR-o models, $RAE_\alpha$ was set to 0.5 Bohr. This approach significantly reduced the theoretical combinatorial search space from 2500 to 121 candidates, with potential for further refinement through targeted optimization strategies. Subsequently, we conducted feature selection before training the RR-i model. To assess the FI, we employed GBR, followed by an iterative last-place elimination procedure guided by the $RAE_{0.5}$ as the evaluation metric. As shown in **Figure S2,** the model achieved a peak average $RAE_{0.5}$ of 0.810 on the test set with 16 features, while the highest single-training $RAE_{0.5}$ reached 0.88. After finalizing the setup, six algorithms were evaluated for the RR-i task. Optimal hyperparameters were selected via grid search and 5-fold cross-validation, with $RAE_{0.5}$ as the performance metric. As summarized in **Table 2**, the GBR-based model outperforms others. Visualization of predictions for 131 test samples (**Figure 3**) reveals larger errors ($\Delta r_{in}$) when the true $r_{in}$ exceeds 3.0 Bohr, whereas predictions are more accurate for $r_{in}$ < 3.0 Bohr. This discrepancy likely stems from the scarcity of high-$r_{in}$ samples, hindering model training. Notably, despite our limited high-$r_{in}$ data, GBR exhibits significant errors only when $r_{in}$ > 3.5 Bohr (a very rare situation nevertheless). Next, the GBR-predicted inner radii were incorporated as inputs to the RR-o model (17 features in total). Following the same workflow as RR-i, all six algorithms were evaluated (see **Table 2**). With predicted $r_{in}$ as inputs, the SVR model achieved the highest $RAE_{0.5}$ of 0.90 (118 correct predictions out of 131 test samples). Visualization of RR-o predictions (**Figure 4**) indicates slightly degraded performance in high $r_{out}$ regions due to data sparsity, while accuracy remains robust in low-radius regions.

In conclusion, we proposed a metal/insulator classification model based on a new supervised machine learning framework, which involves the structural information besides the common chemical composition. With the consideration of the practical crystal structure, the prediction accuracy has increased from 92% to 95.2%. For non-metals, the exact band gap prediction could be carried out using the shDFT-1/2 method, which involves optimizing two parameters $r_{in}$ and $r_{out}$ jointly. We have developed a regression strategy to predict $r_{in}$ and $r_{out}$ based on the chemical composition and structural information of binary compounds. This could typically reduce the number of trials from 2500 to 121, while the accuracy is still kept tolerable. Specifically, the absolute prediction error is generally less than 0.2 Bohr for small radius regime (within 3 Bohr). By significantly reducing the computational cost of shDFT-1/2 calculations, our method provides an efficient yet accurate solution for high-throughput band gap predictions in materials genomics—particularly for complex systems involving defects, interfaces, or other non-ideal structures. Future work may extend this framework to multicomponent compounds, and the model precision may be further refined.

**Acknowledgement**

This work was supported by the National Natural Science Foundation of China under Grant No. 12474230.

Supplementary Information for

# Acceleration of shell DFT-1/2 in high-throughput calculations via cutoff radii prediction

Shanzhong Xie, Kan-Hao Xue,* Zijian Zhou, Xiangshui Miao

School of Integrated Circuits, Huazhong University of Science and Technology, Wuhan 430074, China

*Corresponding author, email: xkh@hust.edu.cn

## Supplementary Note 1

This study involves two datasets, denoted as D1 (DATASET 1, used for training the MC model) and D2 (DATASET 2, used for training the RR model).

Dataset D1 consists of data corresponding to 11805 binary compounds, among which 9173 are metallic and 2632 are non-metallic. It contains 54 initial features, as detailed in **Table S1**. These include 16 elemental descriptors (yielding 32 features for binary compounds), 6 structural descriptors, and 16 derived features calculated from common ratios or differences between elemental descriptors. The classification labels in D1, indicating whether a compound is metallic or non-metallic, are sourced from the Materials Project t[1]. Compounds with a reported band gap $E_g$ = 0 are labeled as metallic; otherwise, they are labeled as non-metallic.

Dataset D2 comprises 653 samples and includes 68 initial features, as listed in **Table S2**. These features consist of 18 elemental descriptors (resulting in 36 for binary compounds), 8 structural descriptors, and 24 derived features based on elemental descriptor ratios or differences. The labels in D2 correspond to two cutoff radii, determined through a comprehensive scanning process that ideally involves a two-dimensional (2D) joint optimization of $r_{in}$ and $r_{out}$.

**Table S1** The 54 initial features of D1 and their descriptions.

| Features | Description |
|---|---|
| $Z_A$, $Z_B$, $Z_{A-B}^{diff}$ | Atomic numbers of sites A and B and their difference |
| $A_A$, $A_B$, $A_{A-B}^{diff}$ | Atomic masses of sites A and B and their difference |
| $\chi_{Pual}^A$, $\chi_{Pual}^B$, $\chi_{Pual}^{A-B}$ | Pauling electronegativities of sites A and B and their difference |
| $r_{calculated}^A$, $r_{calculated}^B$, $r_{calculated}^{A/B}$ | Calculated atomic radii of sites A and B and their ratio |
| $r_{covalent}^A$, $r_{covalent}^B$, $r_{covalent}^{A/B}$ | covalent radii of sites A and B and their ratio |
| $IP_1^A$, $IP_1^B$, $IP_1^{A-B}$ | First ionization potentials of sites A and B and their difference |
| $V_E^A$, $V_E^B$, $V_E^{A-B}$ | Valence electron counts of sites A and B and their difference |
| $EA^A$, $EA^B$, $EA^{A-B}$ | Electron affinities of sites A and B and their difference |
| $G^A$, $G^B$, $G^{A-B}$ | Group numbers of sites A and B and their difference |
| $P^A$, $P^B$, $P^{A-B}$ | Period numbers of sites A and B and their difference |
| $\alpha^A$, $\alpha^B$, $\alpha^{A/B}$ | Atomic polarizabilities of sites A and B and their ratio |
| $\sigma^A$, $\sigma^B$ | Electrical conductivities of sites A and B |
| $T_{boiling}^A$, $T_{boiling}^B$, $T_{boiling}^{A-B}$ | Boiling points of sites A and B and their difference |
| $T_{melting}^A$, $T_{melting}^B$, $T_{melting}^{A-B}$ | Melting points of sites A and B and their difference |
| $\rho^A$, $\rho^B$, $\rho^{A/B}$ | Densities of sites A and B and their ratio |
| $\lambda^A$, $\lambda^B$ | Thermal conductivities of sites A and B |
| $C_{max}^A$, $C_{min}^A$, $C_{max-min}^A$ | Maximum and minimum coordination numbers of site A and their difference |
| $r_{max}$, $r_{min}$, $r_{max-min}$ | Maximum (farthest atom) and minimum (nearest |

|  | neighbor) coordination distances and their difference |
|---|---|
| $SG$ | Space group number. |
| $ca_{ratio}$ | Ratio of the unit cell's c-axis to a-axis lattice parameters |

**Table S2** The 68 initial features of D2 and their descriptions.

| Features | Description |
|---|---|
| $Z_A$, $Z_B$, $Z_{A-B}^{diff}$ | Atomic numbers of sites A and B and their difference |
| $A_A$, $A_B$, $A_{A-B}^{diff}$ | Atomic masses of sites A and B and their difference |
| $\chi_{Pualing}^{A}$, $\chi_{Pualing}^{B}$, $\chi_{Pualing}^{A-B}$ | Pauling electronegativities of sites A and B and their difference |
| $r_{calculated}^{A}$, $r_{calculated}^{B}$, $r_{calculated}^{A/B}$, $r_{calculated}^{A+B}$, $r_{calculated}^{A-B}$ | Calculated atomic radii of sites A and B, their ratio, sum, and difference |
| $r_{covalent}^{A}$, $r_{covalent}^{B}$, $r_{covalent}^{A/B}$, $r_{covalent}^{A+B}$, $r_{covalent}^{A-B}$ | Single-bond covalent radii of sites A and B, their ratio, sum, and difference |
| $IP_1^A$, $IP_1^B$, $IP_1^{A-B}$ | First ionization potentials of sites A and B and their difference |
| $V_E^A$, $V_E^B$, $V_E^{A-B}$ | Valence electron counts of sites A and B and their difference |
| $EA^A$, $EA^B$, $EA^{A-B}$ | Electron affinities of sites A and B and their difference |
| $G^A$, $G^B$, $G^{A-B}$ | Group numbers of sites A and B and their difference |
| $P^A$, $P^B$, $P^{A-B}$ | Period numbers of sites A and B and their difference |
| $\alpha^A$, $\alpha^B$, $\alpha^{A/B}$ | Atomic polarizabilities of sites A and B and their ratio |
| $\sigma^A$, $\sigma^B$ | Electrical conductivities of sites A and B |
| $T_{boiling}^{A}$, $T_{boiling}^{B}$, $T_{boiling}^{A-B}$ | Boiling points of sites A and B and their difference |
| $T_{melting}^{A}$, $T_{melting}^{B}$, $T_{melting}^{A-B}$ | Melting points of sites A and B and their difference |
| $\rho^A$, $\rho^B$, $\rho^{A/B}$ | Densities of sites A and B and their ratio |
| $\lambda^A$, $\lambda^B$ | Thermal conductivities of sites A and B |
| $C_{max}^{A}$, $C_{min}^{A}$, $C_{max-min}^{A}$ | Maximum and minimum coordination numbers of site A and their difference |
| $r_{max}$, $r_{min}$, $r_{max-min}$ | Maximum (farthest atom) and minimum (nearest neighbor) coordination distances and their difference |
| $SG$ | Space group number. |
| $ca_{ratio}$ | Ratio of the unit cell's c-axis to a-axis lattice parameters |
| $r_{s+p}^{A}$, $r_{s+p}^{B}$, $r_{s+p}^{A/B}$, $r_{s+p}^{A+B}$, $r_{s+p}^{A-B}$ | Combined s-orbital and p-orbital radii of sites A and B, their ratio, sum, and difference |
| $\chi_{MB}^{A}$, $\chi_{MB}^{B}$, $\chi_{MB}^{A-B}$ | Mulliken-Berkowitz electronegativities of sites A and B and their difference |
| $\xi_{metal}$ | Metallic property indicator (1 for metallic, 0 for non-metallic) |
| $\kappa_{ion}$ | Ratio of cation count to anion count within the unit cell |

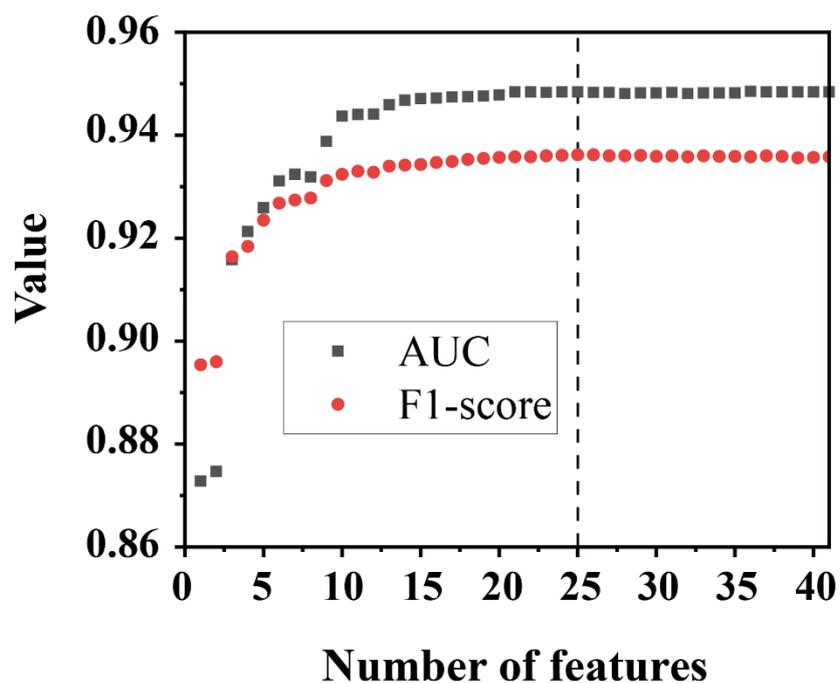

**Figure S1**. The feature selection process using the "last-place elimination" methodology in training the MC model. The plot shows the relationship between the number of features selected and their corresponding AUC and F1-score values. The vertical dashed line represents the optimal number of features for the best model performance.

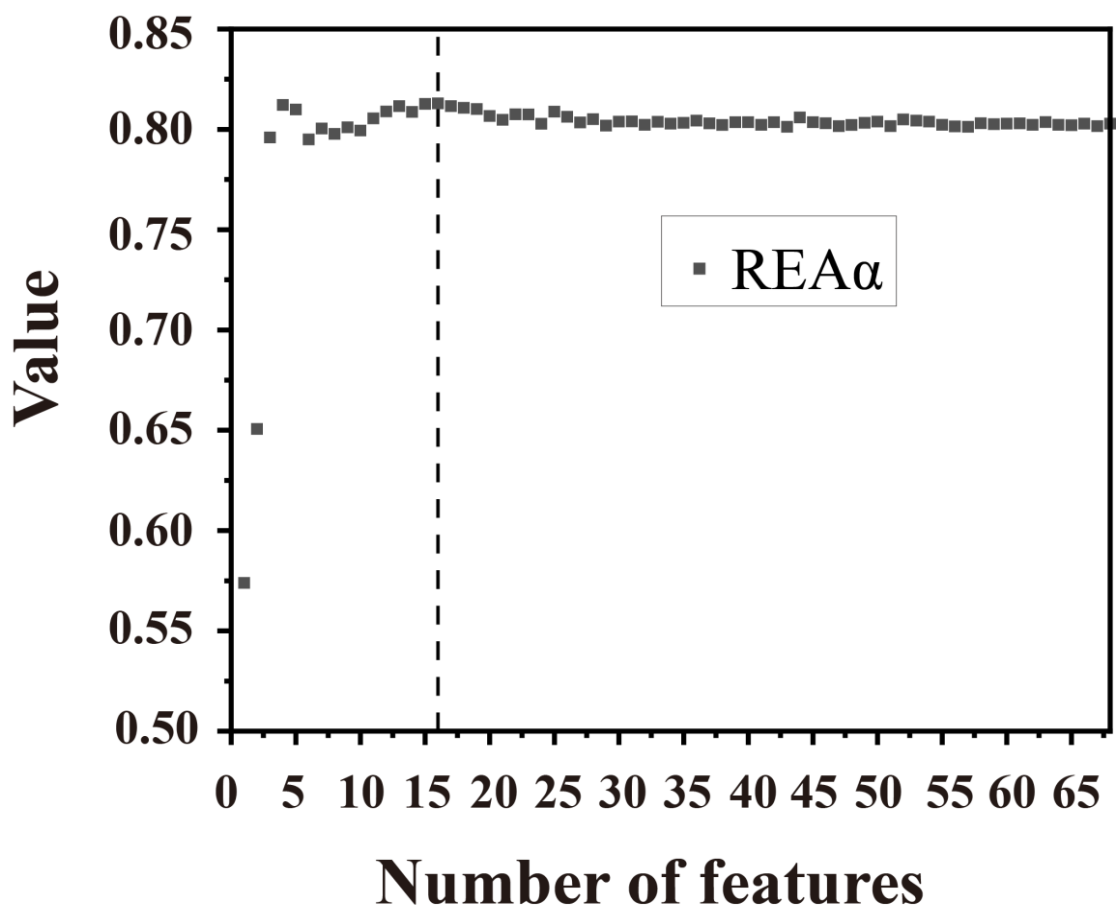

**Figure S2.** The feature selection process using the "last-place elimination" methodology in training the RR model. The plot shows the relationship between the number of features selected and their corresponding $RAE_\alpha$. The vertical dashed line represents the optimal number of features for the best model performance.